\documentstyle[preprint,aps]{revtex}
\tightenlines

\begin{document}
\newcommand{\beq}{\begin{equation}}
\newcommand{\eeq}{\end{equation}}
\newcommand{\beqa}{\begin{eqnarray}}
\newcommand{\eeqa}{\end{eqnarray}}
\newcommand{\sr}{\sqrt}
\newcommand{\fr}{\frac}
\newcommand{\mn}{\mu \nu}
\newcommand{\G}{\Gamma}

\draft
\preprint{ INJE-TP-00-08, hep-th/0012082 }
\title{Mass generation with Gauss-Bonnet
term : No van Dam-Veltman-Zakharov discontinuity  in AdS space}
\author{  Y.S. Myung\footnote{E-mail address:
ysmyung@physics.inje.ac.kr} }
\address{
Department of Physics, Graduate School, Inje University,
Kimhae 621-749, Korea}
\maketitle

\begin{abstract}
We prove  that in anti de Sitter space,
there is no van Dam-Veltman-Zakharov discontinuity
in the  graviton propagator.
Here we obtain the mass term of $M^2\propto \Lambda^2$  from the
Gauss-Bonnet term, which is a ghost-free one.
The condition that the massless limit is smooth is automatically
satisfied for this case.

\end{abstract}

\newpage

Recently there has been much interest in the massless limit of the
massive graviton propagator\footnote{For a spin-3/2 supersymmetric
particle,
its correct massless limit is not $m \to 0$\cite{GNie} but $m \to \sqrt{- \Lambda/3}$
\cite{DW}. I thank Waldron for pointing out this.}\cite{KMP,POR,GNie}.
A key word of this approach is that van  Dam-Veltman-Zakharov (vDVZ)
discontinuity\cite{DVZ}
is peculiar to Minkowski space and it does not arise in anti de Sitter space.
They used the spin-2 Pauli-Fierz mass term\cite{FP} for this calculation.
As is well known, there is no origin for this in the starting
action. A criterion for introducing the Pauli-Fierz term is that
it is a  ghost-free term for a free, massive spin-2 propagation in
the linearized level\footnote{However, if one considers the massive graviton from the
Kaluza-Klein reduction, the Pauli-Fierz term could be expected to appear.
 I thank Papazoglou for pointing out this point.}. We need another ghost-free term which can be
derived from the starting action.
A concrete example is the Gauss-Bonnet term\cite{ZZZ}, which is obviously a ghost-free
combination. This can generate a mass term in anti de Sitter space.

In this letter, we investigate the the vDVZ discontinuity in anti de
Sitter space using the Gauss-Bonnet term.
For simplicity, we choose the harmonic gauge (transverse gauge :$\nabla_\mu h^{\mn}=
\fr{1}{2} \nabla^\nu h$).
This choice is obvious for the massless propagation because this
has the gauge symmetry. Although the gauge symmetry is broken in the massive case,
we can use this as the transversal  condition to study the massive propagation.

We start from the 4D gravity with a cosmological constant and the
Gauss-Bonnet term\cite{ZZZ}

\beq
I = \int d^4x \sqrt{- g} \Big \{ \fr{1}{16 \pi G} R -\Lambda +
\alpha( R^2 -4 R^2_{\mn} + R^2_{\mu\nu\rho\sigma})\Big \}.
\label{Action}
\eeq
Here we set $16 \pi G=2$ and $\Lambda<0$.
Also we follow $\eta_{\mn} = {\rm diag}(-+++)$ convention.
The equation of motion is given
by
\beqa
\mbox{} &&R_{\mn}-\fr{1}{2}g_{\mn}(R- 2\Lambda)=
g_{\mn}\alpha( R^2 -4 R^2_{\mn} + R^2_{\mu\nu\rho\sigma})
\nonumber  \\
\mbox{} && ~~~~-2\alpha(2RR_{\mu\nu}-4R_{\mu\rho}R_{\nu}~^\rho
 -4R_{\rho\mu\sigma\nu}R^{\rho\sigma} +
 2R_{\mu\rho\sigma\kappa}R_{\nu}~^{\rho\sigma\kappa}).
\label{FEQ}
\eeqa
The  anti de Sitter space solution is expressed in terms of the metric ($\bar g_{\mn})$  as
\beq
\bar R_{\mu\nu\rho\sigma}=\fr{\Lambda}{3}(\bar g_{\mu\rho}\bar g_{\nu\sigma}-
\bar g_{\mu\sigma}\bar g_{\nu\rho}),~~~
\bar R_{\mn}=\Lambda \bar g_{\mn},~~~ \bar R=4 \Lambda.
\label{Sol}
\eeq

To study the propagation of the metric, let us introduce the perturbation
around the background space
\beq
g_{\mn} = \bar g_{\mn} +h_{\mn}.
\label{Per}
\eeq
Hereafter we use  the background values  without "overbar" (for example,
$ \bar g_{\mn} \to g_{\mn}$). Further we use the gauge condition for a simple
calculation.
After a lengthy calculation, its linearized equation to
Eq.(\ref{FEQ}) with the external source $T_{\mn}$ takes the form
\beq
 \Delta_L h_{\mn} +\fr{1}{2}\nabla^2h g_{\mn}  - 2\Lambda( h_{\mn}
 -\fr{1}{2}h g_{\mn}) + \fr{64\Lambda^2 \alpha}{9}(h_{\mn}-\fr{1}{4}h g_{\mn}) =4 T_{\mn}
\label{PEQ}
\eeq
with the Lichnerowicz operator $\Delta_L^{(2)}$ for spin-2 field.
In deriving this , we use the relations
\begin{eqnarray}
&& 2 \delta R_{\mu\nu\alpha\beta}(h)= -\nabla_\alpha\nabla_\mu h_{\nu\beta}
-\nabla_\beta\nabla_\nu h_{\mu\alpha}+ \nabla_\beta\nabla_\mu h_{\nu\alpha}
+\nabla_\alpha\nabla_\nu h_{\mu\beta}
\nonumber \\
&& ~~~~+ R_{\mu\gamma\alpha\beta} h^\gamma~_\nu
+ R_{\nu\gamma\alpha\beta} h^\gamma~_\mu,
\label{eqc} \\
&& 2\delta R_{\mn}(h)=\Delta_L^{(2)} h_{\mn}= -\nabla^2h_{\mn}
-2 R_{\rho\mu\alpha\nu} h^{\rho\alpha} + 2R_{\rho(\mu}h^\rho~_{\nu)},
\label{eqr} \\
&& \delta R(h)= g^{\mu}\delta R_{\mn}(h)- h^{\mu}R_{\mn}
\label{eqs}
\end{eqnarray}
Here we observe that the Gauss-Bonnet term contributes to the
linearized equation as a traceless combination.
The trace of Eq.(\ref{PEQ}) takes the form
\beq
 \nabla^2h +2 \Lambda h = 4 T.
\label{tre}
\eeq
Using this, Eq.(\ref{FEQ}) leads to
\beq
 \Delta_L h_{\mn}  -2 \Lambda h_{\mn}  +
 M^2_{GB}( h_{\mn}-\fr{1}{4}h g_{\mn}) =4 T_{\mn}-2 T
 g_{\mn},
\label{PEQf}
\eeq
where the Gauss-Bonnet mass of the graviton
is determined  as $M_{GB}^2=\fr{64\Lambda^2 \alpha}{9} $.
Now we are interested in the transverse traceless metric
propagation ($h_{\mn}^{tt}$) in anti de Sitter space. This corresponds to a true
graviton propagation.  Let us compare Eq.(\ref{PEQf}) with
Eq.(3) in ref.\cite{POR}
\beq
 \Delta_L h_{\mn} + 2\nabla_{(\mu} \nabla^\rho h_{\nu)\rho}
 -\nabla_{(\mu} \nabla_{\nu)}h  -2 \Lambda h_{\mn}  +
 M^2_{PF}(h_{\mn}+ \fr{1}{2}h g_{\mn}) =4 T_{\mn}-2 T
 g_{\mn},
\label{PO}
\eeq
 Under the harmonic gauge, two equations are the same forms
except the trace term (=$\cdots h g_{\mn}$). However this term
is irrelevant to our purpose because we consider  the transverse traceless sector.
In order to study this propagation, we take
\beq
 \Delta_L h_{\mn}^{tt}  -2 \Lambda h_{\mn}^{tt}  +
   M_{GB}^2 h_{\mn}^{tt} =4 T_{\mn}^{tt}.
\label{PEQtt}
\eeq
Here the transverse traceless source ($ \nabla^\mu T_{\mn}^{tt}=0,
T^{tt}=0$) is given by\cite{POR}
\beq
T_{\mn}^{tt}= T_{\mn} - \fr{1}{3} T g_{\mn} +\fr{1}{3}(\nabla_\mu
\nabla_\nu + g_{\mn}\Lambda/3)(\nabla^2 + 4 \Lambda/3)^{-1}T.
\eeq
According to Porrati in\cite{POR}, the result is clear.
The propagator with both $\Lambda$ and $M_{GB}^2$ has only a physical
pole at $\nabla^2= M_{GB}^2 -2\Lambda$. Its residue is given by
\beq
2T_{\mn}T^{\mn} + T \Big[ \fr{2 \Lambda -2M_{GB}^2}{3M_{GB}^2-2
\Lambda} \Big] T.
\eeq
For $\Lambda \to 0$ at the fixed $M_{GB}^2(>\Lambda,\alpha \to$ large), we have
one-particle amplitude for the massive spin-2 field as
\beq
A_{massive}= \fr{2}{-\nabla^2 +M_{GB}^2} \Big( T^{\mn}T_{\mn}-
\fr{1}{3} T^2 \Big).
\eeq
This is positive (the ghost is free)\cite{MKL}.
On the other hand, for $ M_{GB}^2(<\Lambda,\alpha \to$ small)$\to 0$ at finite $\Lambda$,
 one finds
the massless spin-2 amplitude
\beq
A_{massless}= \fr{2}{-\nabla^2 } \Big( T^{\mn}T_{\mn}-
\fr{1}{2} T^2 \Big)
\eeq
which leads also to a positive one.
Concerning the smooth massless limit of $M^2/\Lambda \to
0$\cite{KR}, our case provides a good limit. This is so because
$M^2_{GB}=\fr{64\Lambda^2 \alpha}{9} $ vanishes faster than the
cosmological constant $\Lambda$, as $\Lambda \to 0$. For this case
we assume that the Gauss-Bonnet parameter $\alpha$
is small and finite. If we include the Pauli-Fierz mass term instead of the
Gauss-Bonnet term of $M^2_{GB}(h_{\mn}-h g_{\mn}/4)$,
this contributes   $M^2_{PF}(h_{\mn}-h g_{\mn})$ to
Eq.(\ref{PEQ}). For comparison, we note the contribution of the cosmological
term as $-2\Lambda(h_{\mn}-h g_{\mn}/2)$ with $\Lambda<0$. All of these  play the same role
of the mass-like parameter in anti de Sitter space.

At this stage we wish to comment on the origin of our mass.
Miemiec\cite{MI} showed that the lowest mass of  a localized graviton in 4D anti de
Sitter space gets a quadratic one of $\Lambda$. That is, $M^2 \propto
|\Lambda^2|$. Also Schwartz\cite{SCH} argued that in addition to
 a bare mass ($M^2 \propto |\Lambda|$) of graviton, there also
 exists a CFT correction to the graviton propagator  via the AdS/CFT correspondence. This
 takes a form of $\delta M^2 \propto |\Lambda|^2$.
 We note that the Gauss-Bonnet term may arise  from the
   quantum correction of the matters in the curved space.
   Hence our mass
 term which is proportional to $\Lambda^2$ is not an unphysical one but it
   may give us a new evidence for
   the AdS/CFT correspondence.

In conclusion, we obtain the mass term from the Gauss-Bonnet term.
This is a ghost- free combination. Importantly we generate the mass
term  through the starting action.
Although the trace part of its linearized equation differs slightly from
the Pauli-Fierz mass term, the transverse traceless part is
exactly the same as Eq.(3) in ref.\cite{POR}. Hence we find that
there is no the vDVZ discontinuity in the graviton propagator.

\section*{Acknowledgments}

I would like to thank  H. W. Lee  for helpful
discussions.
This work was supported by the Brain Korea 21 Program, Ministry of
Education, Project No. D-0025.


\begin{thebibliography}{99}
\bibitem{KMP} I. I. Kogan, S. Mouslopoulos, and A. Papazolou,
hep-th/0011138.

\bibitem{POR} M. Porrati, hep-th/0011152.

\bibitem{GNie} P. A. Grassi, P. van Nieuwenhuizen, hep-th/0011278.
\bibitem{DW} S. Deser and A. Waldron, hep-th/0012014.

\bibitem{DVZ} H. van Dam and M. Veltman, Nucl Phys. {\bf B22},
397(1970); V. I. Zakharov, JETP Lett. {\bf 12}, 312(1970).

\bibitem{FP} M. Fierz, Helv. Phys. Acta. {\bf 12}, 3(1939); M.
Fierz and W. Pauli, Proc. Roy. Soc. {\bf 173}, 211(1939).


\bibitem{ZZZ} D. G. Boulware and S. Deser, Phys. Rev. Lett. {\bf 55}, 2656(1985);
             B. Zwiebach, Phys. Lett. {\bf B156}, 315(1985);
             B. Zumino, Phys. Pept. {\bf 137}, 109(1986);
             I. Low and A. Zee, hep-th/0004124; J. E. Kim and H.
             M. Lee, hep-th/0010093;
             N. E. Mavromatos and J. Rizos, hep-th/0008074.

\bibitem{MKL} Y.S. Myung, G. Kang, and H.W. Lee, Phys. Lett.
{\bf B478}, 294 (2000), hep-th/0001107;G. Kang and Y.S. Myung, Phys. Lett. {\bf B483}, 235
(2000), hep-th/0003162; Y.S. Myung and G. Kang, hep-th/0005206; Y. S. Myung, hep-th/0009117.

\bibitem{KR} A. Karch and L. Randall, hep-th/0011156; I. I. Kogan,
S. Mouslopoulos, and Papazoglou, hep-th/0011141.

\bibitem{MI} A. Miemiec, hep-th/0011160.
\bibitem{SCH}  M. D. Schwartz, hep-th/0011177.


\end{thebibliography}
\end{document}